\documentclass{iopjournal}
\usepackage{amsmath}
\usepackage{graphicx,epsfig}
\usepackage{subcaption}
\usepackage{microtype}
\usepackage{ragged2e}
\justifying

\begin{document}
\articletype{paper}
\title{Relativistic calculations of electron impact excitation cross-sections of neutral tungsten}
\author{Ritu Dey$^{1,*}$\orcid{0000-0001-7665-0497}, Ayushi Agarwal$^{2,**}$\orcid {}, Reetesh Kumar Gangwar$^1$\orcid {0000-0002-6376-8012}, Deepti Sharma$^3$\orcid{0009-0007-6962-2683}, M. B. Chowdhuri$^{3,4}$\orcid{0000-0001-8313-1089}, Rajesh Srivastava$^2$\orcid{0000-0001-5632-8070}, and Joydeep Ghosh$^{3,4}$\orcid{0000-0002-4369-1900}}

\affil{$^1$ Department of Physics, Indian Institute of Technology Tirupati, Yerpedu 517619, India}\\
\affil{$^2$ Department of Physics, Indian Institute of Technology Roorkee, Roorkee 247667, India}\\
\affil{$^3$ Institute for Plasma Research, Gandhinagar 382428, India}\\
\affil{$^4$ Homi Bhabha National Institute, Training School Complex, Anushakti Nagar, Mumbai 400094, India}\\
\affil{$^*$Author to whom any correspondence should be addressed.\\
       $^{**}$Presently working at University of Luxembourg, Belval, Luxembourg}

\email{deyritu1@gmail.com}

\begin{abstract}
\justifying
Reliable tungsten spectroscopy is essential for modelling and diagnosing tungsten-containing plasmas, particularly in edge/divertor conditions where neutral tungsten (W I) is produced by sputtering and contributes to visible/near-UV emission. Quantitative interpretation of such spectra requires internally consistent atomic structure (level energies, radiative rates) and electron-impact excitation (EIE) data as inputs to collisional-radiative models. In this work, we present fine-structure-resolved EIE cross sections for W I computed using the relativistic distorted-wave (RDW) method. Target states are described by multi-configurational Dirac-Fock (MCDF) wavefunctions with an extensive configuration-interaction (CI) expansion including valence-valence and core-valence correlation through hole and deeper-core configurations. The resulting energies are benchmarked against available theoretical results and the recommended values from the NIST database. Fine-structure-resolved cross sections are reported for excitations from the ground level 5d$^4$6s$^2$ ($^5D_0$) and six metastable levels 5d$^4$6s$^2$ ($^5D_1$, $^5D_2$, $^5D_3$, $^5D_4$, $^3P_0$) and 5d$^5$($^6S$)6s ($^7S_3$) into excited levels belonging primarily to the 5d$^4$6s($^6D$)6p and 5d$^5$($^6S$)6p configurations, over incident electron energies from threshold to 500 eV. With the exception of a limited set of previously studied transitions, most of the reported fine-structure-resolved cross sections are presented here for the first time. The results show a strong dependence on the initial level and indicate that excitation from the  metastable state yields the largest cross sections among the states considered, highlighting the importance of metastable populations in W I modelling. In addition, radiative transition probabilities for selected prominent transitions are calculated and compared with existing data. The present dataset is expected to be valuable for collisional-radiative modelling and spectroscopic diagnostics of tungsten plasmas in the $\sim$ 1-50 eV range.

\end{abstract}
\section{Introduction}
Tungsten is a key plasma-facing material in magnetic-confinement fusion because it combines a very high melting point with favourable thermo-mechanical properties, relatively low erosion under many operating conditions, and low tritium retention. For these reasons, tungsten is widely used for divertor and wall components and will remain central to the plasma-surface interaction in present devices and in ITER-class reactors \cite{pitts}. At the same time, tungsten is a high-Z element with 74 electrons and therefore a very efficient radiator once it enters the confined plasma. Even trace tungsten concentrations can lead to substantial radiative power losses and associated plasma cooling, which directly impacts confinement, operational margins, and plasma control. Consequently, a quantitative understanding of tungsten source generation at the wall, impurity transport through the edge, and radiative behaviour across charge states is essential for both predictive modelling and diagnostics-driven operation. \par
In hot core plasmas, highly charged tungsten ions dominate radiative losses in the VUV/EUV to soft X-ray spectral bands, and a substantial literature addresses atomic data and spectroscopy of those charge states\cite{pitts,putterich,lin,zhang,morita,oishi,alex,rneu}. In contrast, the neutral and low-charge stages are most relevant to the edge/divertor region and to the near-surface source physics: sputtered tungsten enters the plasma initially as neutral W I and, depending on conditions, may remain neutral over appreciable path lengths before ionizing. Moreover, visible/near-UV W I lines are frequently observed and exploited as indicators of tungsten influx and erosion \cite{beigmann, rneu1,vanrooji,nishijima,cshin}. These neutral lines are also important in laboratory plasmas and in devices where tungsten is used as a diagnostic emitter. For this reason, the atomic kinetics of W I are a necessary component of tungsten impurity modelling at the boundary and an important element of tungsten spectroscopy as a routine diagnostic.\par
Interpreting tungsten emission spectroscopy in a quantitatively meaningful way requires collisional-radiative modelling (CRM) that links plasma parameters to level populations and line emissivities through a network of radiative and collisional processes. In this framework, atomic structure and collision data determine how tungsten responds to a given plasma environment. For neutral tungsten, the essential atomic inputs include the fine-structure level energies and their ordering, transition wavelengths and radiative rates ($A$-values) together with electron-impact excitation (EIE) cross sections or rate coefficients connecting the ground and metastable states to the excited configurations that radiate most strongly. The importance of these datasets is not merely formal: the widely used S/XB concept for estimating impurity influx and erosion \cite{cshin}, for instance, depends explicitly on excitation rates and on the probability that excitation leads to photon emission in a measured line versus being diverted through competing radiative or collisional pathways. Similarly, CRMs require reliable excitation data because level populations are typically far from Boltzmann equilibrium in low-density edge plasmas; metastable populations, cascading, and radiative trapping are often decisive in line intensities. Therefore, missing or inconsistent excitation cross sections and radiative parameters directly influence into systematic errors in inferred tungsten influx and in simulated spectral emissivities.\par

Neutral tungsten is particularly challenging from the perspective of atomic physics because of its open $5d$ shell and strong relativistic effects. The ground configuration $[Xe]4f^{14}5d^46s^2$ produces a dense manifold of fine-structure levels with strong configuration interaction and substantial spin-orbit mixing. Due to this complexity, level energies and term assignments can be sensitive to the configuration-interaction (CI) expansion and to the treatment of electron correlation (including core-valence correlation involving the $5p$ shell), making it non-trivial to obtain a target structure that is simultaneously accurate and computationally tractable. Second, strong mixing can activate excitation channels that would be nominally “forbidden” under strict LS-coupling selection rules, thereby redistributing excitation strength among many closely spaced levels. Third, because many low-lying levels are long lived, metastables can accumulate significant populations under typical edge/divertor conditions, and EIE from metastable initial levels can become as important as (or more important than) excitation from the ground level. Each of these features complicates both the construction of atomic data and its use in modelling: if a dataset is limited to a handful of transitions from the ground state, or if it omits key metastable channels, it will generally be insufficient for predictive CRM applications.\par
In view of these challenges, we focus this work on calculation of atomic structure for W I that yields reasonable fine-structure energies and wavefunctions for the levels of interest, especially for the low-lying metastables and for the excited configurations that dominate visible/near-UV emission. The second is a corresponding set of EIE cross sections computed with a scattering method consistent with that structure description, covering the dominant excitation pathways and spanning an incident electron-energy range suitable for edge and laboratory plasmas. Existing work \cite{smyth,duck1} has advanced both components, but the overall picture remains incomplete: while high-charge tungsten ions have been explored from extensive works \cite{tdas1, dipti1, dipti2, priti2, dipti3, priti3, shukla1, shukla2, goyal, mccann} , fine-structure-resolved W I collision datasets remain comparatively sparse, often limited to selected lines or restricted energy intervals \cite{smyth,duck1,deypfr}. In practical modelling, tungsten diagnostics frequently rely on a small set of observed lines, but those lines are fed by a much larger network of excitation and cascade channels. Incomplete excitation data therefore compromises not only absolute emissivities but also line ratios, metastable sensitivity, and the reliability of inferred plasma parameters.\par
The present paper addresses this need by providing an expanded and internally consistent dataset of W I atomic structure and EIE cross sections suitable for implementation in CRMs and for diagnostic interpretation. We compute fine-structure level structures using a multi-configurational Dirac-Fock (MCDF) approach with extensive CI, and we then calculate EIE cross sections using the relativistic distorted-wave (RDW) method implemented in the Flexible Atomic Code (FAC) \cite{gu}. The MCDF+CI target description is chosen because relativistic effects and strong configuration mixing are essential for W I, while FAC provides a practical and widely used platform to generate large amounts of consistent structure and collision data. Using this framework, we focus on excitations from the ground level $5d^46s^2$ ($^5$D$_0$) and from a set of long-lived low-lying metastable levels, namely $5d^46s^2$ ($^5$D$_{1-4}$, $^3P_0$) and  5d$^5$($^6$S)6s ($^7S_3$) into excited levels primarily belonging to the $5d^46s^2$ and  5d$^5$($^6$S)6p configurations. These excited configurations are central because they contain many of the strong visible/near-UV transitions frequently exploited in tungsten diagnostics. The EIE calculations are performed from threshold up to incident electron energies of 500 eV, providing coverage that is relevant to a broad range of plasma conditions and that facilitates the construction of Maxwellian-averaged rate coefficients over edge-relevant temperature windows.\par
A distinguishing aspect of this work is the explicit treatment of multiple initial metastable levels, rather than restricting attention to ground-state excitation alone. This point is crucial for W I because the low-lying fine-structure manifold supports metastables with lifetimes long enough to accumulate under typical boundary plasma conditions. In CR modelling, metastables act as alternative ``starting points'' for excitation, often with substantially different coupling strengths and accessibility to final states. Consequently, the effective excitation rates into diagnostically important upper levels can be strongly metastable-dependent. Our calculations quantify this dependence by providing cross sections from each selected initial state and by comparing their magnitudes and energy dependence across transitions. In particular, we find that excitation from 5d$^5$($^6$S)6s ($^7S_3$) yields the largest cross sections among the initial states considered for many of the important channels, while the $5d^46s^2$ quintet metastables provide additional strong pathways that can reshape predicted populations. This implies that diagnostic modelling that ignores metastables can misestimate emissivities even when the observed lines are from ground because the population flows into upper manifolds can be dominated by metastable excitation and subsequent cascade feeding. Another motivation for presenting atomic structure results alongside the EIE data is that, in heavy neutral systems such as W I, uncertainties in level energies and mixing propagate directly into collision strengths and threshold behaviour. Near-threshold cross sections are especially sensitive to excitation energies, and the correct ordering and spacing of low-lying levels matter when computing excitation into closely spaced fine-structure manifolds. For this reason, we benchmark the calculated energy levels against recommended datasets (such as those compiled in NIST \cite{Kramida}) and against available theoretical calculations \cite{smyth,duck1}, and we discuss the extent to which the present CI model reproduces the key low-lying structure relevant for the subsequent EIE calculations. In addition to collision data, we compute radiative quantities ($A$ values) for selected transitions.\par
The RDW approach is  a balanced approcah between physical realism and tractability for a complex heavy neutral. Close-coupling methods (such as R-matrix) are capable of describing resonance structures near threshold that can strongly enhance effective collision strengths at low temperatures. However, such calculations become extremely demanding for W I because of the enormous target complexity and the dense excitation spectrum. Distorted-wave methods, while generally less complete in their treatment of near-threshold resonances, remain a practical tool for producing broad excitation datasets across many levels and a wide energy range, particularly when the emphasis is on providing extensive coverage needed for CR modelling rather than resolving resonance fine structure for a small subset of transitions. In this work, we therefore employ RDW to generate fine-structure-resolved cross sections for a large set of transitions, and we include comparisons to available literature data for selected lines to assess the overall level of agreement and to indicate where resonance physics may become important.\par
The resulting dataset is intended to support tungsten plasma diagnostics and modelling in regimes where neutral tungsten is present and where visible/near-UV emission is used for inference. In edge and divertor plasmas, uncertainties in EIE cross sections and radiative branching ratios directly affect predicted emissivities and inferred tungsten influx via S/XB-type approaches. More generally, reliable W I data improve CRMs that couple impurity transport to radiation, and they enable more consistent interpretation of tungsten line emission in both steady-state and transient plasma conditions. By reporting cross sections from the ground and key metastables to the primary $6p$ excited configurations, and by providing supporting structure and radiative data in a consistent relativistic framework, this work aims to reduce one of the persistent atomic-data bottlenecks for W I modelling. \par
The remainder of the paper is organized as follows. Section 2 outlines the theoretical basis for
the relativistic distorted wave approach. Section 3 presents energy levels and electron
impact excitation cross sections for transitions in neutral W, with a focus on fine-structural resolution. The radiative transition probabilities are also provided for prominent neutral W emission spectra. A concluding remark is
given in Section 4.

\section{Theory}
RDW theory within FAC is employed in this work to calculate electron impact excitation cross-sections from ground and metastable states. The transition matrix element (T) with initial state (a) and final state (b) of the target atom is expressed as \cite{ayushi},

\begin{eqnarray}
 T_{a\rightarrow b}^{RDW}(J_{a(b)},M_{a(b)},{\bf{k_{a(b)}}},\mu_{a,(b)},\theta)=
 \langle \Phi_b^{rel}(1,2,...,N)F_{b, \mu_b}^{DW-}\\ \nonumber (k_b,N+1)|V-U|A\Phi_a^{rel}(1,2,...,N)F_{a, \mu_a}^{DW+} (k_a,N+1)\rangle
 \end{eqnarray}
 where, $F_a^{DW-}$ and $F_b^{DW+}$ are the relativistic distorted wave function of the incoming and outgoing projectile electron, respectively, with the wave vector ${\bf{k_{a(b)}}}$ and spin component. $\Phi_a$ and $\Phi_b$ are the relativistic bound state wavefunction with the N electrons of the target atom, respectively. In Eq. (1), $J_a$, $M_a$ are the total angular momentum and magnetic quantum number in the initial bound state whreas, $J_b$, $M_b$ are the same quantities in the final bound state after the excitation process. ``$A$'' is the anti-symmetrization operator ensures the total wavefunction obeys Fermi-Dirac antisymmetry. Within the relativistic distorted wave approximation, the incident and scattered electrons are described by wavefunctions distorted by the electrostatic potential of $U(r)$. The distortion potential U(r) is typically the spherically averaged static potential of the atom. V is the interaction potential between target atom and the projectile electron as follows:

 \begin{equation}
V=-\frac{Z}{r_{N+1}}+\sum_{i=1} ^{N} \frac{1}{r_i-r_{N+1}},
 \end{equation}
$r_i$ (i = 1, 2,..., N)  is the position vectors of the i$^th$ bound electron in the target atom, each measured from the nucleus of the atom. So, if the target has N electrons, there are N such vectors. $r_{N+1}$  denotes the position of the incoming (projectile) electron, also measured relative to the same nucleus. The subscript ``N+1'' is used because the total number of electrons in the system (target  and projectile). The distortion potential is expressesd as \cite{priti3}:
\begin{equation}
U({r_{N+1}})=-\frac{Z}{N+1}+\sum_{q \in all~ subshells} N_q \int_{0}^{\infty} \left[ P^2_{n_q \kappa_q} (r) + Q^2_{n_q \kappa_q} (r)\right]\frac{1}{r_>},
 \end{equation}
 where, $N_q$ is the occupation number of the q$^{th}$ subshell, and the electron in it is represented by quantum numbers $n_q$, $\kappa_q$. $P$ and $Q$ represent the larger and smaller components of the radial wavefunctions of atomic orbitals. In Eq. (1), $J_a$, $M_a$ are the total angular momentum and magnetic quantum number in the initial bound state whreas, $J_b$, $M_b$ are the same quantities in the final bound state after the excitation process. ``$A$'' is the anti-symmetrization operator ensures the total wavefunction obeys Fermi-Dirac antisymmetry. $F_a^{DW-}$ and $F_b^{DW+}$ which are the relativistic distorted wave function of the incoming and outgoing projectile electron, respectively, expressed as follows:

\begin{eqnarray}
 F^{DW \pm}_{a(b),\mu_{a(b)}}({\bf{k_{a(b)}}},N+1)=\frac{1}{(2\pi)^{3/2}}\left[ \frac{E_{a(b)}+c^2}{2E_{a(b)}}\right]^{1/2}\sum_{\kappa m}4\pi i^l e^{\pm i\eta_{\kappa}}\\ \nonumber
 a_{a(b),\kappa m}^{\mu_{a(b)}}\widehat{(k_{a(b)})}\frac{1}{r}
  \left( \begin{array}{cc}
f_{\kappa}(r) & \chi_{\kappa m}(\hat{r}) \\
 ig_{\kappa}(r) & \chi_{-\kappa m}(\hat{r}) \end{array} \right).
 \end{eqnarray}
Where, $\eta_{\kappa}$ is the phase shift of the partial wave, $f_{\kappa}$ and $g_{\kappa}$ are the large and small components
of the radial wavefunctions, and $\chi_{\pm \kappa m}$ is the spinor spherical harmonic, $k$ and $\mu$ are the momentum, and spin projections of the projectile electron, respectively. E is the relativistic energy and is expresses as: $E_{a(b)}=\sqrt{k_{a(b)}^2 c^2+c^4}$. Finally, the excitation cross-section is expressed as:

\begin{equation}
 \sigma_{ab}=(2\pi)^4 \frac{k_b}{2(2J_a+1)k_a} \int  |T_{a \rightarrow b}^{RDW}|^2.
 \end{equation}
 Since the relativistic distorted wave theory is inherently perturbative in nature, its accuracy is limited in the near-threshold (low-energy) region. Therefore, a correction factor is introduced to improve the reliability of the cross sections at low incident electron energies. The corrected cross-section, as a function of the incident electron energy, is expressed as \cite{wunderlich}:
\begin{eqnarray}
\sigma_{{ab}{[corrected]}}=[1-(E_{threshold}/E)^{\frac{1}{2}}]\sigma_{ab}.
 \end{eqnarray}
\noindent
Here, $E$ corresponds to the incident electron energy, while
$E_{threshold}$ corresponds to the excitation threshold for the transition from level $a$ to $b$. The excitation threshold energy, $E_{thresold}$, is obtained from the energy difference, $E_j-E_i$ where, $E_i$ and $E_j$ are the initial and final state energies listed in Table~\ref{Table1} of the following section.

\section{Results and Discussions:}
The fine-structure spectrum of neutral tungsten considered in the present calculations is based on a large configuration expansion. That model includes a total of 43,556 fine-structure resolved energy levels arising from extensive excited-state configurations. The configurations that are taken for the present calculations are as follows: 5p$^6$5d$^4$6[s$^2$,p$^2$,d$^2$], 5p$^6$5d$^4$6s6[p,d], 5p$^6$5d$^4$6s7[s,p,d,f], 5p$^6$5d$^6$, 5p$^6$5d$^3$6s6[p$^2$,d$^2$], 5p$^6$5d$^2$6s$^2$6p$^2$, 5p$^6$5d$^2$6s$^2$6d$^2$, 5p$^6$5d$^4$7s$^2$, 5p$^6$5d$^3$6s7s$^2$, 5p$^6$5d$^2$6s$^2$7s$^2$, 5p$^6$5d$^3$6s$^2$6d, 5p$^6$5d$^5$7s, 5p$^6$5d$^3$6s$^2$7s, 5p$^6$5d$^5$6[s,p,d], 5p$^6$5d$^3$6s$^2$6p, 5p$^6$5d$^3$6p$^3$, 5p$^6$5d$^2$6s6p$^3$, 5p$^6$5d$^4$6s8[s,p,d,f], 5p$^6$5d$^4$6s9[s,p,d,f], 5p$^6$5d$^3$6s$^2$7p, 5p$^6$5d$^5$7p, 5p$^6$5d$^4$6s5f, 5p$^6$5d$^5$5f, 5p$^6$5d$^3$6d$^3$, 5p$^6$5d6s$^2$6d$^3$, 5p$^5$5d$^5$6s6p, 5p$^5$5d$^6$6p, 5p$^5$5d$^6$6s, 5p$^5$5d$^7$, 5p$^5$5d$^5$6s$^2$, 5p$^5$5d$^5$6s7[s,p,d], 5p$^5$5d$^5$6s8[s,p,d], 5p$^5$5d$^6$6p, 5p$^5$5d$^4$6s$^2$6p, 5p$^4$5d$^6$6s$^2$, 5p$^4$5d$^7$6s, 5p$^4$5d$^8$, 5p$^4$5d$^6$6s6p, 5p$^4$5d$^7$6p, 5p$^4$5d$^5$6s$^2$6p, 5p$^4$5d$^6$6s$^2$, 5p$^4$5d$^7$6s, 5p$^4$5d$^8$. Due to these large configurations, FAC calculations are parallelized and performed on 32 CPU cores of ANTYA, the IPR Linux cluster, to obtain EIE cross-sections. The configurations and energy levels from FAC are listed in Table~\ref{Table1}. This table also contains the energy levels from NIST \cite{Kramida} as well as other approaches reported by Duck-Hee Kwon et al \cite{duck1}. Compared with the NIST energies, the present energy values show good overall agreement, particularly for the low-lying levels, where the deviations remain relatively small. For several excited configurations, the discrepancies increase. Similarly, the GRASP \cite{smyth} calculation shows a maximum 39\% deviation from the NIST energy values. It is worth mentioning that, further inclusions of states, there is no significant improvement in energy levels. Finally, energy levels are corrected to the corresponding NIST \cite{Kramida} values to calculate the EIE cross-sections.
\par
\begin{figure}

     \begin{subfigure}[ht]{0.5\textwidth}
           \includegraphics[width=\textwidth]{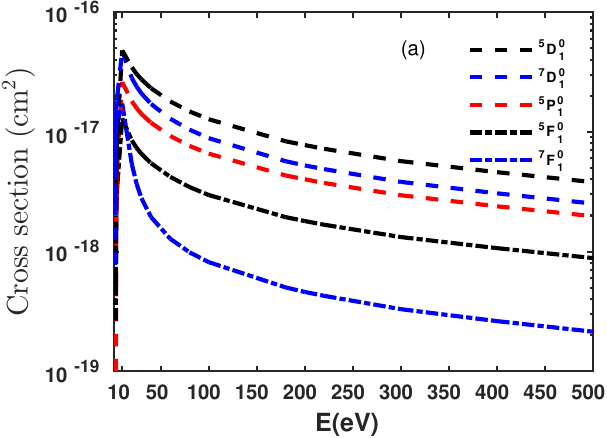}
     \end{subfigure}
     \hfill
     \begin{subfigure}[ht]{0.5\textwidth}
            \includegraphics[width=\textwidth]{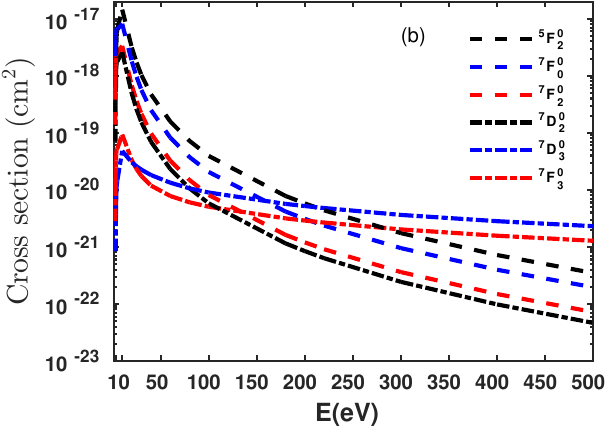}
     \end{subfigure}
\hfill
     \begin{subfigure}[ht]{0.5\textwidth}
            \includegraphics[width=\textwidth]{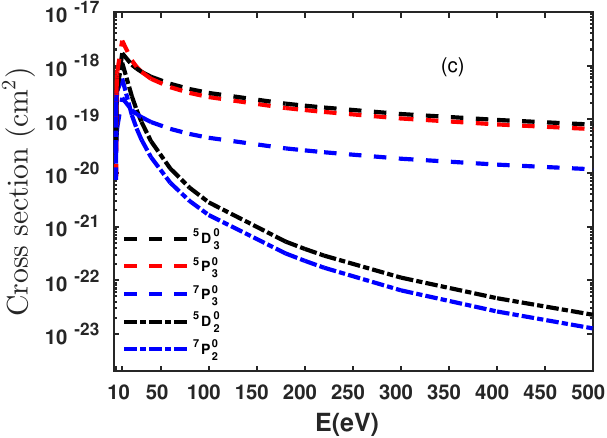}
     \end{subfigure}
\hfill
     \begin{subfigure}[ht]{0.5\textwidth}
            \includegraphics[width=\textwidth]{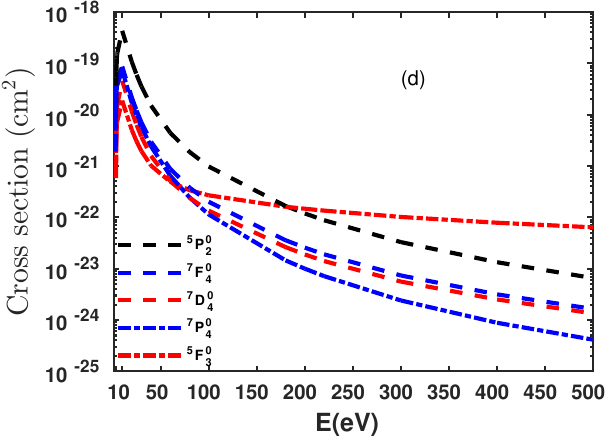}
     \end{subfigure}

      \caption{Electron impact excitation cross-sections of W I from ground state 5d$^4$6s$^2$ ($^5$D$_0$) to various excited states (a-d).}

\label{Figure1}
\end{figure}
\begin{figure}

     \begin{subfigure}[ht]{0.5\textwidth}
           \includegraphics[width=\textwidth]{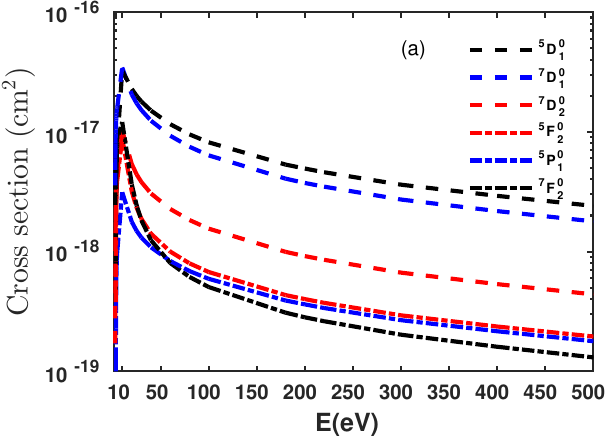}
     \end{subfigure}
     \hfill
     \begin{subfigure}[ht]{0.5\textwidth}
            \includegraphics[width=\textwidth]{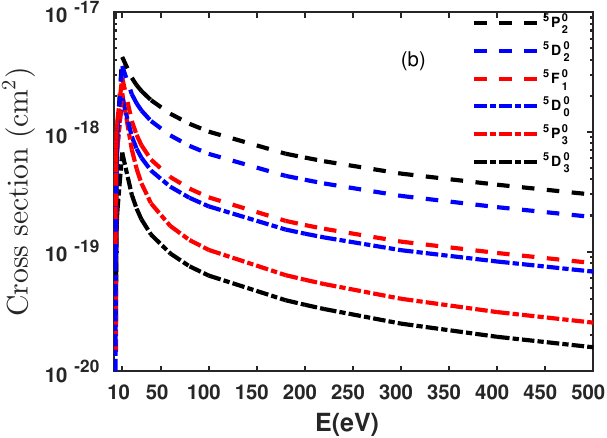}
     \end{subfigure}
\hfill
     \begin{subfigure}[ht]{0.5\textwidth}
           \includegraphics[width=\textwidth]{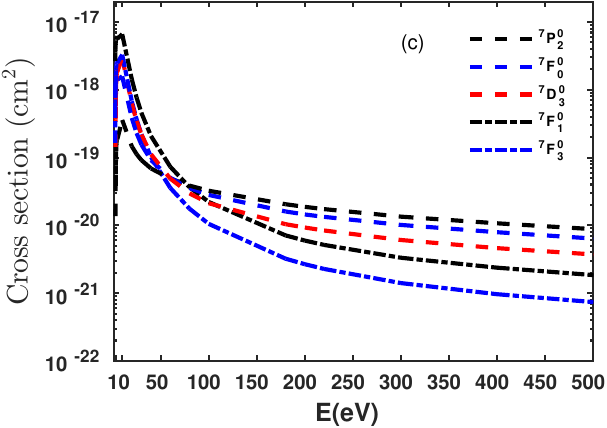}
     \end{subfigure}
\hfill
     \begin{subfigure}[ht]{0.5\textwidth}
           \includegraphics[width=\textwidth]{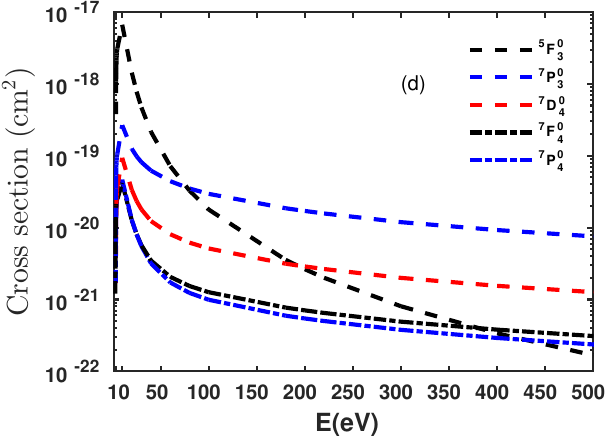}
     \end{subfigure}

      \caption{Electron impact excitation cross-sections of W I from metastable state 5d$^4$6s$^2$ ($^5$D$_1$) to various excited states (a-d), respectively.}
\label{Figure2}

\end{figure}

In the RDW framework, electron-impact excitation cross sections are governed primarily by parity and angular-momentum selection rules. When $\Delta J=0,\pm1$ ($0 \not\leftrightarrow 0$) and the initial and final states have opposite parity, the direct (dipole-allowed $E1$) part of the scattering amplitude contributes, leading to larger cross sections that decrease slowly with increasing electron energy $E$. When this $E1$ coupling is absent (forbidden $\Delta J$ or same parity), excitation occurs through exchange and higher-multipole interactions, which result in smaller peak magnitudes and a steeper fall-off at high energy. Consequently, for both allowed and forbidden transitions, $\sigma(E)$ rises sharply near threshold and decreases gradually with increasing $E$.  The simulations are carried out from the ground state and six long-lived metastable states of W~I into the fine-structure levels of the $5d^46s(^6D)6p$ and $5d^5(^6S)6p$ configurations, for incident electron energies up to 500~eV. Figs.~\ref{Figure1}-\ref{Figure6} provide an overview of these results: Figs.~\ref{Figure1}(a)-(d) show excitation cross-sections from the $5d^46s^2,{}^5D_0$ ground state. Similarly, detailed cross-section results for the other metastable initial states ($5d^5(^6S)6s,{}^7S_3$, $5d^46s^2,{}^5D_{1,2,3,4}$, and $5d^46s^2,{}^3P_0$) are provided in Figs.~\ref{Figure2}-\ref{Figure6}, respectively. In the following, the excitation cross sections from each transition from the ground and metastable states of W I have been discussed.

\begin{table}[htbp]
\centering
\caption{Configurations and upper energy levels of W I transitions from FAC, HFR \cite{hfr}, GRASP \cite {smyth}, MDFGME \cite {duck1} and the NIST \cite{Kramida} recommended values based on experiment.}

\resizebox{\textwidth}{!}{%

\begin{tabular}{lllllllll}
\hline
No. & Level & \multicolumn{7}{l}{Energy (eV)} \\
    &       & FAC & NIST & HFR & GRASP & MDFGME$_1$ & MDFGME$_2$ & MDFGME$_3$ \\
\hline
1 & $5d^46s^2\ (^5D_0)$ & 0.00000 & 0.00000 & 0.00000 & 0.00000 & 0.00000 & 0.00000 & 0.00000 \\
2 & 5d$^4$6s$^2$ ($^5D_1$) & 0.16454 & 0.20709 & 0.22130 & 0.12735 & 0.14639 & 0.15387 & 0.17717 \\

3 & 5d$^5(^6S)$6s ($^7S_3$) & 0.15951 & 0.36591 & 0.36958 & 0.40587 & 0.0748 & 0.15495 & 0.34489\\

4 & 5d$^4$6s$^2$ ($^5D_2$) & 0.35509 & 0.41231 & 0.43009 & 0.29280 & 0.32367 & 0.33375 & 0.36664 \\

5 & 5d$^4$6s$^2$ ($^5D_3$) & 0.54942 & 0.59884 & 0.61073 & 0.47308 & 0.54982 & 0.54943 & 0.64061 \\

6 & 5d$^4$6s$^2$ ($^5D_4$) & 0.74887 & 0.77110 & 0.76942 & 0.66343 & 0.69367 & 0.70692 & 0.74369 \\

7 & 5d$^4$6s$^2$ ($^3P_0$) & 1.26816 & 1.18133 & 1.19170 & 1.39285 & 1.37648 & 1.35536 & 1.27808 \\

8 & 5d$^4$6s($^6D$)6p ($^7F_0^{\circ}$) & 1.39638 & 2.40398 & 2.41469 & 2.11152 & 2.09097 & 2.58115 & 2.49842 \\

9 & 5d$^4$6s($^6D$)6p ($^7F_1^{\circ}$) & 1.47580 & 2.48766 & 2.50344 & 2.18512 & 2.16591 & 2.65574 & 2.57997 \\

10 & 5d$^4$6s($^6D$)6p  ($^7F_2^{\circ}$)& 1.61919 & 2.65931 & 2.66145 & 2.31955 & 2.2991 & 2.78887 & 2.72272\\

11 & 5d$^4$6s($^6D$)6p ($^7F_3^{\circ}$) & 1.80328 & 2.85750 & 2.85530 & 2.50105 & 2.47404 & 2.96369 & 2.90491\\

12 & 5d$^4$6s($^6D$)6p ($^7F_4^{\circ}$) & 2.01916 & 3.07027 & 3.07438 & 2.72406 & 2.68417 & 3.17366 & 3.11951 \\
13 & 5d$^4$6s($^6D$)6p ($^7F_5^o$) & 2.24793 & 3.30746 & 3.00176 & 2.93583 & 3.42526 & 3.37372 & 0.6 \\

14 & 5d$^4$6s($^6D$)6p ($^7D_1^{\circ}$) &1.71201 & 2.65995 & 2.67250 & 2.43030 & 2.49521 & 2.98051 & 2.87602 \\

15 & 5d$^4$6s($^6D$)6p ($^7D_2^{\circ}$)& 1.97813 & 2.97124 & 2.98292 & 2.63331 & 2.74053 & 3.22691 & 3.15133 \\

16 & 5d$^4$6s($^6D$)6p ($^7D_3^{\circ}$) & 2.23309 & 3.24705 & 3.26343 & 2.82352 & 2.95241 & 3.44373 & 3.391\\

17 & 5d$^4$6s($^6D$)6p ($^7D_4^{\circ}$) & 2.45182 & 3.57040 & 3.58466 & & 3.27819 & 3.81824 & 3.75869\\

18 & 5d$^4$6s($^6D$)6p ($^5F_1^{\circ}$) & 2.61011 & 3.22156 & 3.25009 & 3.21238 & 3.27574 & 3.83019 & 3.7035\\

19 & 5d$^4$6s($^6D$)6p ($^5F_2^{\circ}$) & 2.31938 & 3.42972 & 3.43357 & 3.36395 & 3.36524 & 3.96014 & 3.84867 \\

20 & 5d$^4$6s($^6D$)6p ($^5F_3^{\circ}$) & 2.44405 & 3.61279 & 3.62321 & & 3.65538 & 4.15418 & 4.04803 \\

21 & 5d$^5$($^6S$)6p ($^7P_2^{\circ}$) & 3.98058 & 3.25208 & 3.25643 & 2.97101 & 2.96929 & 3.54905 & 3.48621  \\

22 & 5d$^5$($^6S$)6p ($^7P_3^{\circ}$) & 4.08854  & 3.40809 & 3.41889 & 3.11605 & 3.11683 & 3.67771 & 3.61534\\

23 & 5d$^5$($^6S$)6p ($^7P_4^{\circ}$) & 4.45841 & 3.45788 & 3.47785 & 2.98067 & 3.11479 & 3.60894 & 3.56685\\

24 & 5d$^4$6s($^6D$)6p ($^5D_0^{\circ}$) & 2.46810 & 3.30163 & 3.30933 & 3.44381 & 3.33893 & 3.85507 & 3.66865\\

25 & 5d$^4$6s($^6D$)6p ($^5D_1^{\circ}$) & 3.93839 & 3.44410 & 3.45043 & 3.53756 & 3.33274 & 3.89912 & 3.78748\\

26 & 5d$^4$6s($^6D$)6p ($^5D_2^{\circ}$) & 3.18660 & 3.61982 & 3.62677 & & 3.44602 & 4.02916 & 3.85789 \\

27 & 5d$^4$6s($^6D$)6p ($^5D_3^{\circ}$) & 2.90356 & 3.70872 & & & & \\

28 & 5d$^4$6s($^6D$)6p ($^5P_1^{\circ}$) &  4.17299 & 3.49622 & 3.49362 & 3.30163 & 3.5016 & 4.01523 & 3.84609\\

29 & 5d$^4$6s($^6D$)6p ($^5P_2^{\circ}$) & 3.72621 & 3.64432 & 3.64581 & 3.49048 & 3.51362 & 4.06788 & 3.94684  \\

30 & 5d$^4$6s($^6D$)6p ($^5P_3^{\circ}$) & 3.37457 & 3.79226 & 3.80527 & 3.72954 & 3.73949 & 4.21803 & 4.12033 \\ \hline

\end{tabular}}
\label{Table1}
\end{table}

\subsection{Excitation from the Ground State ($5d^46s^2,{}^5D_0$)}

From the $^5D_0$ ground term ($J_i=0$), only transitions to odd-parity levels with $J_f=1$ satisfy the $\Delta J=0,\pm1$ dipole selection rule (since $0\to0$ is forbidden). Indeed, as shown in Figure1~(a), the largest cross sections from $^5D_0$ are found for excitations into the $5d^46s6p$ levels $^5D^{\circ}_1$ and $^7D^{\circ}_1$. These two levels have energies 3.444 eV and 2.659 eV, respectively, favorable angular momentum coupling to the $^5D_0$ ground state, leading to very large near-threshold cross sections of the order of $\times$$10^{-17}$~cm$^2$ as shown in Fig.~\ref{Figure1}(a). It is also noticed from Fig.~\ref{Figure1}(a), that the excitaions to level such as $^7F^{\circ}_1$, $^5F^{\circ}_1$ and $^5P^{\circ}_{1}$, the cross-sections are still high at the order of $10^{-17}$~cm$^2$. While weaker than the leading channels, these are still significant, especially in plasmas with broad electron energy distributions. Within this group, there is a clear dependence on $J$. Among the five fine-structure levels of the $^7F^{\circ}$ term, the $J=1$ level ($^7F^{\circ}_1$) exhibits relatively large excitation cross sections and decreases more gradually with increasing incident electron energy. In contrast, the higher-$J$ members, namely $^7F^{\circ}_2$, $^7F^{\circ}_3$, and $^7F^{\circ}_4$, show comparatively weaker cross sections, with $^7F^{\circ}_2$ and $^7F^{\circ}_4$ in particular displaying a more rapid fall-off as the incident energy increases (see Figs.~\ref{Figure1}(b) and (d)). As depicted in Figs.~\ref{Figure1}(c) and (d), the excitation cross-sections to the $5d^5(^6S)6p$ ($^7P^{\circ}_{2,3,4}$) levels attain peak values of $2.5 \times 10^{-19}$ cm$^2$, $5.7 \times 10^{-19}$ cm$^2$, and $7.3 \times 10^{-20}$ cm$^2$ for the $^7P^{\circ}_{2}$, $^7P^{\circ}_{3}$, and $^7P^{\circ}_{4}$ levels, respectively. The cross-sections decrease rapidly with increasing incident electron energy $E$ for the $^7P^{\circ}_{2}$ and $^7P^{\circ}_{4}$ levels, whereas the $^7P^{\circ}_{3}$ (blue dashed line in Fig.~\ref{Figure1}(c)) level exhibits a comparatively more gradual decline.

\subsection{Excitation from the Metastable State $5d^46s^2,{}^5D_1$}

For the first excited quintet level $5d^46s^2,{}^5D_1$ ($J_i=1$), a greater number of strong dipole-allowed transitions appear, since $J_i=1$ permits $\Delta J=0$ as well as $\Delta J=\pm1$ (unlike the $J_i=0$ ground state). As seen in Figs.~\ref{Figure2}(a)-(d), several excitation channels out of $^5D_1$ exhibit very large cross sections. Among these, the transitions to $^7D^{\circ}_1$ and $^5D^{\circ}_1$ peaks on the order of $\times$ $10^{-17}$~cm$^2$ and remain relatively large and decrease slowly with the incident energy as shown in Fig.~\ref{Figure2}(a). In addition, the transitions from $^5D_1$ metastable state to $^7D^{\circ}_2$, and $^7F^{\circ}_2$ have significant cross-sections and vary slowly with incident energy $E$. Similarly to $^5D_0$, the transitions from $^5D_1$ metastable state to $^7P^{\circ}_{2,3,4}$ show lower magnitudes in cross-sections (Figs.~\ref{Figure2}(c) and (d)). In addition, the excitation to $^7P^{\circ}_2$ (black dashed line in Fig.~\ref{Figure2}(c)) and $^7P^{\circ}_3$ (blue dashed line in Fig.~\ref{Figure2}(d)) shows similar cross-sections over the entire energy range $E$. Figs.~\ref{Figure2}(a), and (b) show the excitation to $^5F^{\circ}_2$, $^5D^{\circ}_2$ and $^5P^{\circ}_2$, respectively, having similar cross-sections among the $^5$-multiplet final levels. Among, the $^7F^{\circ}$ levels accessible from $^5D_1$, (Fig.~\ref{Figure2}(c)) the cross sections follow $^7F^{\circ}_1 \gg {}^7F^{\circ}_0$ upto incident energy $E$ upto $\sim$ 50 eV, beyond that energy $^7F^{\circ}_0 \gg {}^7F^{\circ}_1$. On the other hand,$^7F^{\circ}_2 \gg {}^7F^{\circ}_3 \gg {}^7F^{\circ}_4$. This indicates that fine-structure effects still modulate strengths within a given spin multiplet. In comparison with the ground $^5D_0$ level, the transitions originating from the metastable $^5D_1$ state to the $^7P^{\circ}_4$, $^7D^{\circ}_4$, and $^7F^{\circ}_4$ levels exhibit similar peak magnitudes, but decrease more gradually with increasing incident electron energy $E$, as shown in Figs.~\ref{Figure1}(d) and \ref{Figure2}(d), respectively.

\subsection{Excitation from the Metastable State $5d^5(^6S)6s,{}^7S_3$}

\begin{figure}
 \begin{subfigure}[ht]{0.5\textwidth}
           \includegraphics[width=\textwidth]{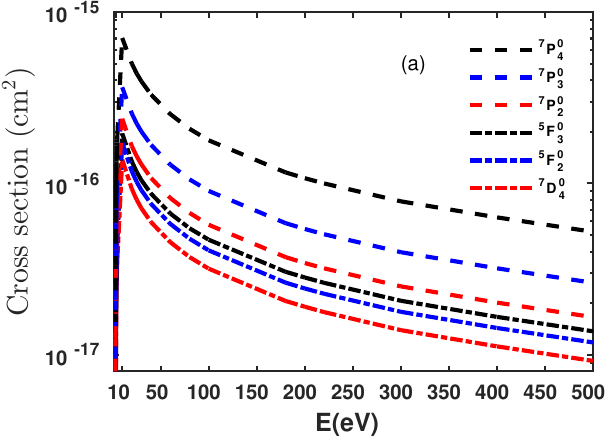}
     \end{subfigure}
     \hfill
     \begin{subfigure}[ht]{0.5\textwidth}
            \includegraphics[width=\textwidth]{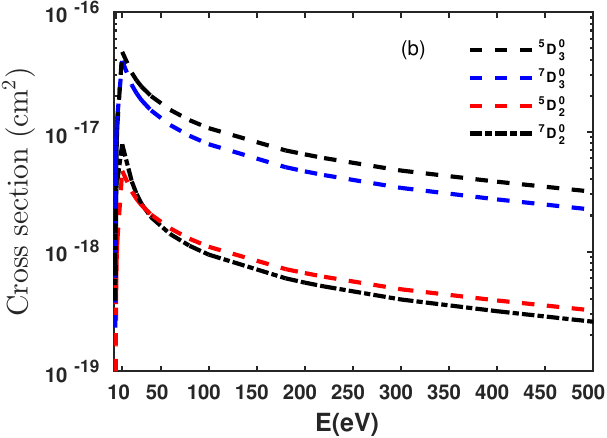}
     \end{subfigure}
\hfill
     \begin{subfigure}[ht]{0.5\textwidth}
           \includegraphics[width=\textwidth]{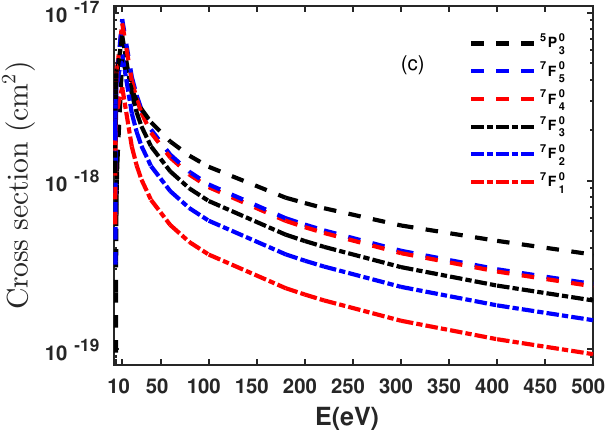}
     \end{subfigure}
\hfill
     \begin{subfigure}[ht]{0.5\textwidth}
           \includegraphics[width=\textwidth]{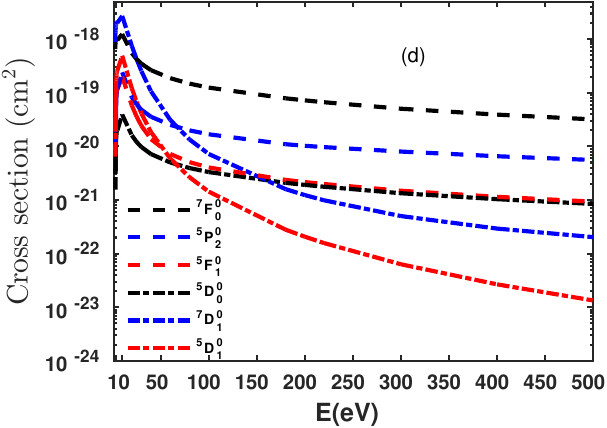}
     \end{subfigure}

\caption{Electron impact excitation cross-sections of W I from metastable state 5d$^5$6s ($^7$S$_3$) to various excited states (a-d).}

\label{Figure3}

\end{figure}

Figs.~\ref{Figure3}(a)-(d) depicted the cross-sections for the transition from the high-spin metastable $5d^5(^6S)6s,{}^7S_3$ ($J_i=3$, $S=3$) to the levels of $5d^46s6p$ and $5d^56p$ ($^7P_{2,3,4}^{\circ}$). These exhibit a somewhat different pattern of excitation, owing to both its larger total spin and its higher angular momentum. With $J_i=3$, a wide set of $\Delta J=0,\pm1$ transitions are dipole-allowed, in fact, nearly all final levels with $J_f=2,3,$ or $4$ can be reached via a significant $E1$ transition. As a result, the group of dominant excitations from $^7S_3$ is broader than in the previous quintet cases. The cross-section to $^7P^{\circ}_4$ transition shows the largest magnitudes at the order of 7 $\times$ 10$^{-16}$ cm$^2$ near threshold and slowly decreases with incident energy as shown in Fig.~\ref{Figure3}(a). This is due to the fact that excitation arises from a strong dipole-allowed valence transition ($\Delta S=0$, $\Delta J=+1$, and even to odd parity transition) with the $5d^5$ core acting as a spectator, resulting in large radial overlap. Among the $^7$-multiplet final states, $^7P^{\circ}_{2,3,4}$, and $^7D^{\circ}_{4}$ dominate over the other states, with a long, flat tail at higher incident energies. In addition, channels $^5F^{\circ}_2$, and $^5F^{\circ}_3$ as shown in Fig.~\ref{Figure3}(a), produce larger cross-sections each an order of magnitude $\sim$ 2 $\times$ 10$^{-16}$ cm$^2$ near threshold. It is also observed that, $^7S_3$ to $^7F^{\circ}$, $^7D^{\circ}$ and $^7P^{\circ}$ excitation cross sections exhibit a clear statistical $(2J_f+1)$ dependence, with higher J fine-structure components showing systematically larger cross sections. Similar behaviour is also noticed for $^7S_3$ to $^5$-multiplet transitions.
Relative to the quintet initial states discussed earlier, the $^7S_3$ metastable shows a somewhat different hierarchy of cross sections. Notably, spin-orbit mixing in tungsten has strongly enhanced several $^5$-multiplet final channels (e.g. $^5F^{\circ}_2$ and $^5F^{\circ}_{3}$), making them important as the $^7$-multiplet channels. Figs.~\ref{Figure3}(b) and (d) which contain $^5D^{\circ}$ channels exhibit larger cross-sections for $^5D^{\circ}_3$ with peak magnitude 4.6$\times$ 10$^{-17}$ cm$^{2}$. As illustrated in Fig.~\ref{Figure3}(d), only transitions to $^5D^{\circ}_0$, $^5D^{\circ}_1$, $^5F^{\circ}_1$, and $^7D^{\circ}_1$ show lower cross-sections, whereas the remaining transitions yield significant excitation cross-sections. Overall, the amplitude distribution for $^7S_3$ is broader than, say, for $^5D_1$. A larger number of channels can be considered as strong, which means the excitation strength from $^7S_3$ is spread over both $^7$ and $^5$ series levels rather than being concentrated in only a few lines. It is to be noted that the transition from $^5D_1$ to $5d^56p$ states, which corresponds to $^7P_{2,3,4}$, has the lower order (10$^3$ or less) of magnitudes with respect to the transition from $^7S_3$.
For collisional-radiative modeling, the $^7S_3$ metastable is expected to be a major feeder into the $^5F^{\circ}_{2,3}$, $^5D^{\circ}_{3}$, $^7P^{\circ}_{2,3,4}$, and $^7D^{\circ}_{4}$ levels, which then cascade radiatively to other configurations. Even a modest population in the $^7S_3$ level could significantly enhance emission from lines associated with the $^5F^{\circ}$ and $^7P^{\circ}$ manifold, given the very large cross sections. Meanwhile, the many weaker channels (e.g. $^7D^{\circ}_{1}$, $^7F^{\circ}_{0}$, $^5F^{\circ}_{1}$, $^5D^{\circ}_{0,1}$,and $^5P^{\circ}_{2}$) shown in Fig.~\ref{Figure3}(d) can likely be omitted or lumped in a simplified kinetics model without losing accuracy, since they contribute less to net excitation rates. The dominant $^5F^{\circ}$ and  $^7P^{\circ}$ pathways, however, must be retained to correctly account for the collisional excitation out of $^7S_3$.


 \begin{figure}

     \begin{subfigure}[ht]{0.5\textwidth}
           \includegraphics[width=\textwidth]{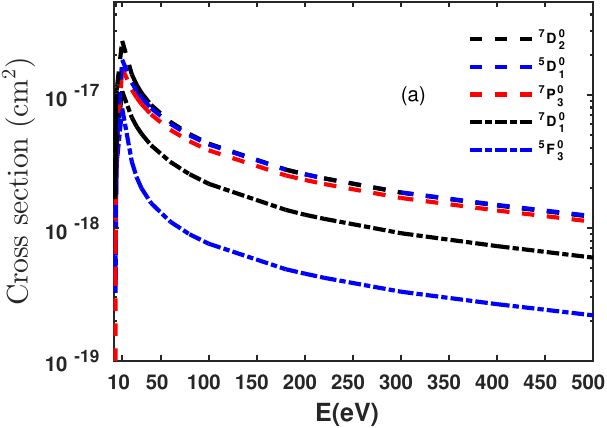}
     \end{subfigure}
     \hfill
     \begin{subfigure}[ht]{0.5\textwidth}
            \includegraphics[width=\textwidth]{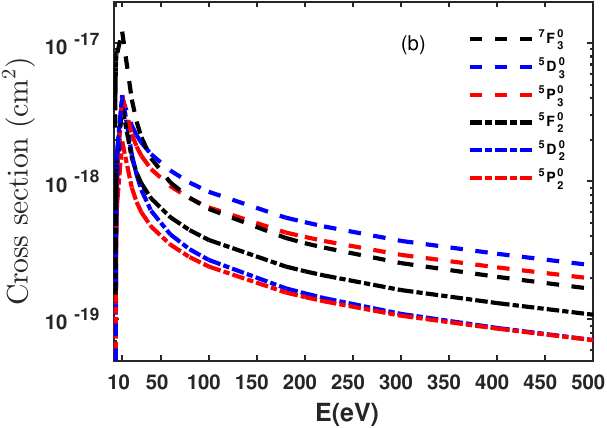}
     \end{subfigure}
\hfill
     \begin{subfigure}[ht]{0.5\textwidth}
           \includegraphics[width=\textwidth]{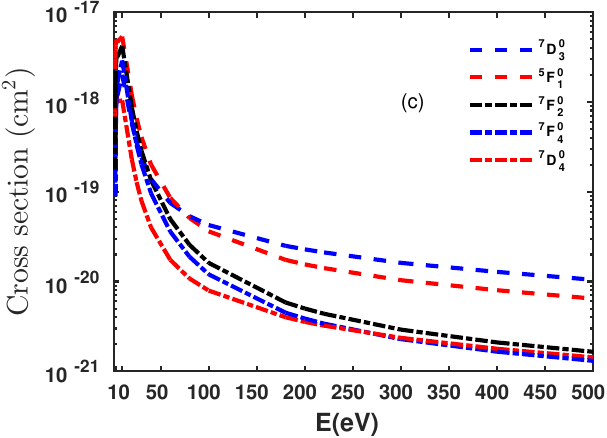}
     \end{subfigure}
\hfill
     \begin{subfigure}[ht]{0.5\textwidth}
           \includegraphics[width=\textwidth]{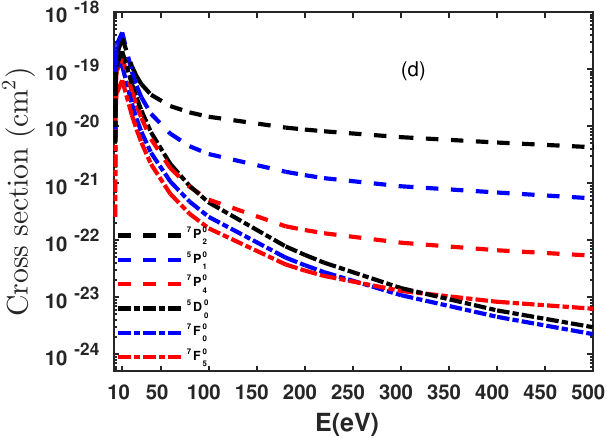}
     \end{subfigure}

\caption{Electron impact excitation cross-sections of W I from metastable state 5d$^4$6s$^2$ ($^5$D$_2$) to various excited states (a-d).}
\label{Figure4}
\end{figure}

\subsection{Excitation from the Metastable State $5d^46s^2,{}^5D_2$}

The metastable $5d^46s^2,{}^5D_2$ ($J_i=2$) yields excitations with magnitudes similar to those of excitations from $5d^46s^2,{}^5D_1$ ($J_i=1$). Figs.~\ref{Figure4}(a)-(d) show that the most prominent transitions from $^5D_2$ are those feeding the $5d^46s6p$, $^5D^{\circ}_1$,  $^7D^{\circ}_{1,2}$, $^7P^{\circ}_{3}$ and $^5F^{\circ}_{3}$ (Fig.~\ref{Figure4}(a)) channels. These channels in particular are the strongest transition among all the cases considered and it peaks of the order of $\sim$ $1-3 \times 10^{-17}$~cm$^2$ and maintains its dominance throughout the entire 0-500~eV range. Figure 4(b) shows the excitation cross sections for the $^7F^{\circ}_{3}$, $^5P^{\circ}_{2,3}$, $^5F^{\circ}_{2}$, and $^5D^{\circ}_{2,3}$ levels. The peak cross sections are of the order of $\sim 4$-$8 \times 10^{-18}$ cm$^{2}$ for most transitions, except for $^7F^{\circ}_{3}$, whose peak cross section is comparatively larger, of the order of $\sim 10^{-17}$ cm$^{2}$, and decreases gradually with increasing incident electron energy $E$. The excitation cross section for the $^7F^{\circ}_{3}$ channel is comparable in magnitude to the dominant channels presented in Fig.~\ref{Figure4}(a). Excitation cross sections to $^5F^{\circ}$ state show a statistical $(2J_{f}+1)$ dependence, with higher J fine-structure components showing systematically larger cross sections ($^5F^{\circ}_3 \gg {}^5F^{\circ}_2 \gg {}^5F^{\circ}_1$) as displayed in Figs.~\ref{Figure4}(a), (b) and (c). As noticed from Fig.~\ref{Figure4}(c), $^7F^{\circ}_{2}$, $^7F^{\circ}_{4}$ and $^7D^{\circ}_{4}$, although peak values are at higher magnitudes, they decrease sharply with incident electron energy $E$. Fig.~\ref{Figure4}(d) displays states like  $^7F^{\circ}_0$, $^7F^{\circ}_5$, and $^5D^{\circ}_0$, more suppressed, with cross sections falling into the $10^{-24}$~cm$^2$ range at the upper end of the energy scale. These weak transitions contribute little and can be safely neglected in most practical analyses, aside from ensuring completeness.

Comparing the $^5D_2$ state with the lower-$J$ cases, the $^5D_2$ metastable is found to produce excitation into the $^7$-multiplet final states that is comparable in strength to that from the $^5D_1$ state. In contrast, when comparing the quintet $^5D_2$ state ($S = 2$) with the septet initial state $^7S_3$ ($S = 3$) discussed above, most excitation channels exhibit somewhat lower cross sections in the $^5D_2$ case.

\subsection{Excitation from the Metastable State $5d^46s^2,{}^5D_3$}

 \begin{figure}
\begin{subfigure}[ht]{0.5\textwidth}
           \includegraphics[width=\textwidth]{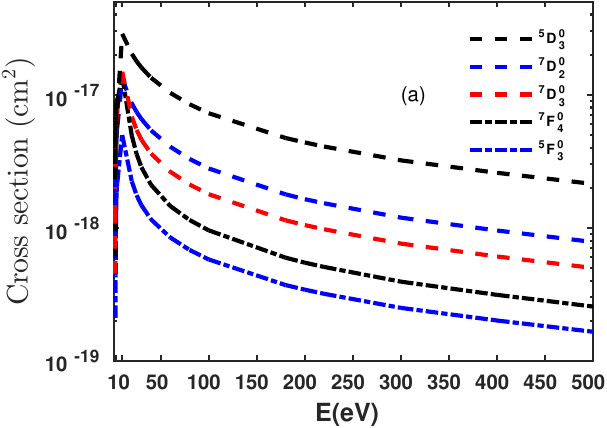}
     \end{subfigure}
     \hfill
     \begin{subfigure}[ht]{0.5\textwidth}
            \includegraphics[width=\textwidth]{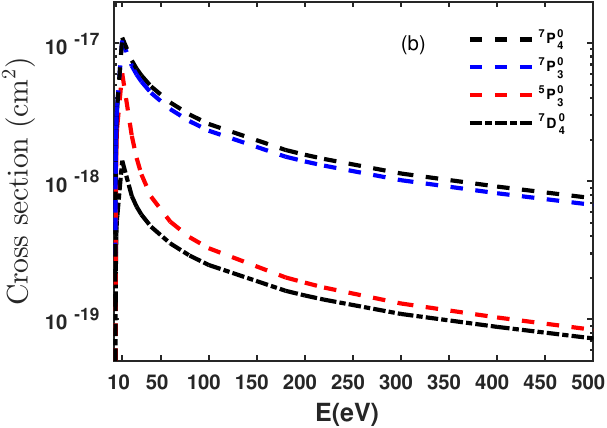}
     \end{subfigure}
\hfill
     \begin{subfigure}[ht]{0.5\textwidth}
           \includegraphics[width=\textwidth]{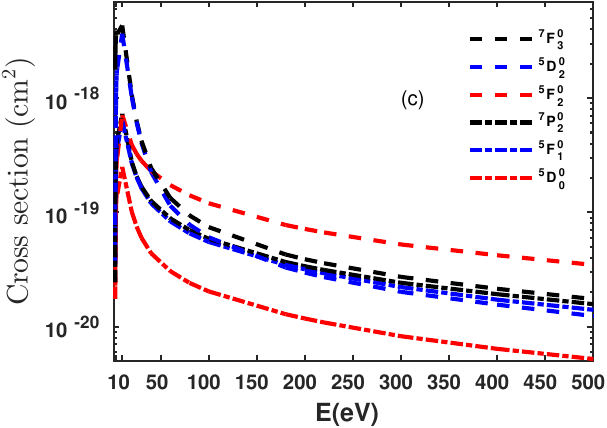}
     \end{subfigure}
\hfill
     \begin{subfigure}[ht]{0.5\textwidth}
           \includegraphics[width=\textwidth]{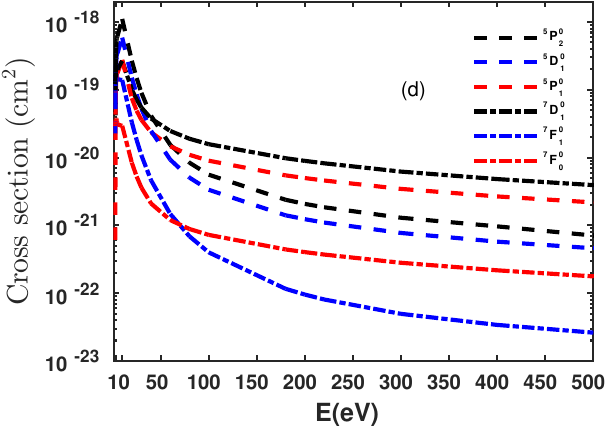}
     \end{subfigure}

\caption{Electron impact excitation cross-sections of W I from metastable state 5d$^4$6s$^2$ ($^5$D$_3$) to various excited states (a-d).}
\label{Figure5}
\end{figure}

For the metastable $5d^46s^2,{}^5D_3$ ($J_i=3$), the overall behavior continues the trends seen for $^5D_2$. As $J_i$ increases, more final levels satisfy $\Delta J=0,\pm 1$, and indeed $^5D_3$ has numerous strong dipole-allowed channels. The most dominant of these is the $^5D^{\circ}_{3}$ channel as shown in Fig.~\ref{Figure5}(a). As illustrated in Fig. \ref{Figure5}(a), the transitions to $^7F^{\circ}_4$ level shows peak cross-section similar as to the transitions to $^7D^{\circ}_{2}$, and $^7D^{\circ}_{3}$. Transitions to $^7P^{\circ}_{3}$, and $^7P^{\circ}_{4}$ in Fig.~ \ref{Figure5}(b) are of comparable order of magnitude, similar to the previous channels $^7D^{\circ}_{2}$, and $^7D^{\circ}_{3}$, exhibiting a gradual decline over the 0-500~eV incident electron energy range. It is also noticed that, $^5P^{\circ}_{3}$ level in Fig.~\ref{Figure5}(b), exhibits cross-sections in the range 6.0$\times$ 10$^{-18}$ cm$^{2}$ to 8.4$\times$ 10$^{-20}$ cm$^{2}$. Fig.~\ref{Figure5}(c) depicted that, channels like $^7F^{\circ}_3$, $^5D^{\circ}_2$, give peak values at the order of $\sim$ 4$\times$ 10$^{-18}$ cm$^{2}$ while decreases very fast with incident electron energy. In addition, the distribution of excitation cross sections among the fine-structure components of the $^7F^{\circ}$ term follows a statistical $(2J_{f}+1)$ dependency, leading to progressively larger cross sections for increasing $J$ except $^7F^{\circ}_0$. Excitation cross-section from $^7F^{\circ}_0$ is higher than the $^7F^{\circ}_1$ beyond the incident electron energy $E \geq 50$ eV, as shown in Fig.~\ref{Figure5}(d). Similar behaviour is also seen for the $^5$-multiplet transitions where excitation cross sections from $^5P^{\circ}_1$ and $^5D^{\circ}_0$ are larger than $^5P^{\circ}_2$ and $^5D^{\circ}_1$ beyond the incident electron energy $\sim$ 50 eV. Interestingly, Fig.~\ref{Figure5}(b) exhibits $^7P^{\circ}_{3}$ and $^7P^{\circ}_{4}$ fine-structure levels with almost identical cross sections across the full incident electron energy $E$.

Comparing the $^5D_3$ results to the $^5D_2$ case, cross-sections are in a similar range. The $^7D_2^{\circ}$ is giving maximum cross-section for $^5D_2$ case, while $^5D_3^{\circ}$ to $^5D_3$ transition is the largest. The main noticeable difference is that both $^7P^{\circ}_2$ and $^7P^{\circ}_3$ are showing similar magnitudes for $^5D_3$, whereas from $^5D_2$ those two transitions have different magnitudes. Comparing $^5D_3$ and $^7S_3$ results, most of the $^7S_3$ transitions give higher excitation cross-sections than $^5D_3$.

From a plasma perspective, the $^5D_3$ metastable state contributes to population in the same way as other quintet metastable states  $^5D_{1}$, and $^5D_{2}$, efficiently feeding a variety of excitation pathways. The dominant channels involve excitation to the $^5D^{\circ}_{3}$, $^7D^{\circ}_{2,3}$, $^7F^{\circ}_{4}$ levels associated with the
$5d^4 6s(^6D)6p$ configuration, along with the $^7P^{\circ}_{3,4}$ levels belonging to the $5d^5(^6S)6p$ configuration. These transitions contribute substantially to the redistribution of population among excited states and are therefore essential for inclusion in comprehensive collisional-radiative models.
\begin{figure}
\begin{subfigure}[ht]{0.5\textwidth}
           \includegraphics[width=\textwidth]{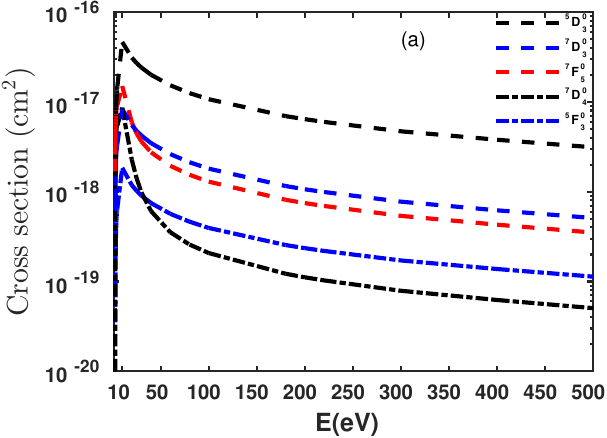}
     \end{subfigure}
     \hfill
     \begin{subfigure}[ht]{0.5\textwidth}
            \includegraphics[width=\textwidth]{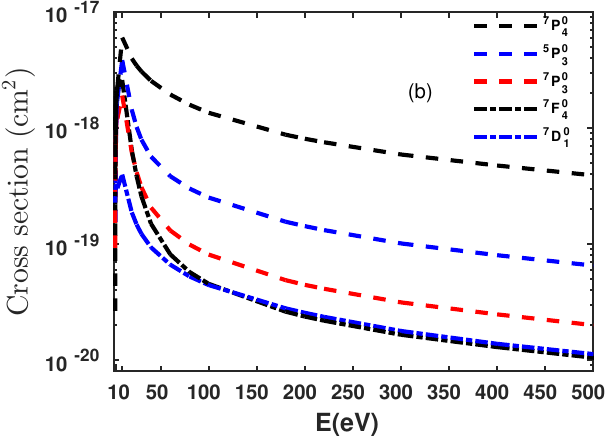}
     \end{subfigure}
\hfill
     \begin{subfigure}[ht]{0.5\textwidth}
           \includegraphics[width=\textwidth]{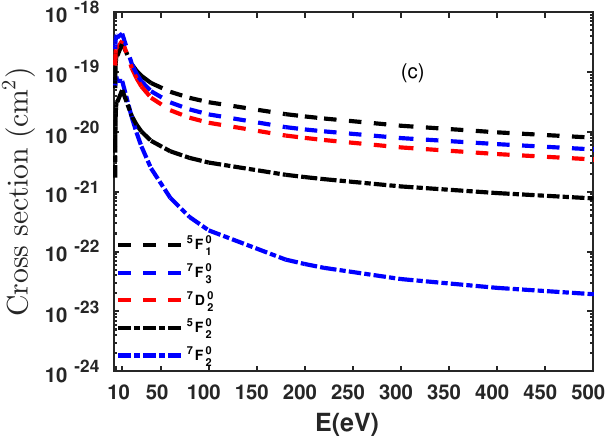}
     \end{subfigure}
\hfill
     \begin{subfigure}[ht]{0.5\textwidth}
           \includegraphics[width=\textwidth]{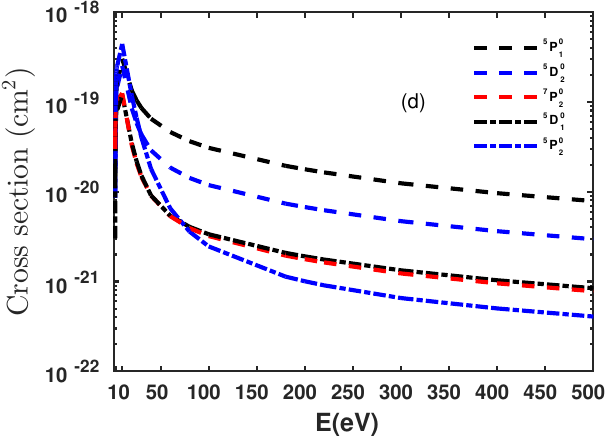}
     \end{subfigure}

\caption{Electron impact excitation cross-sections of W I from metastable state 5d$^4$6s$^2$ ($^5$D$_4$) to various excited states (a-d).}
 \label{Figure6}
\end{figure}

\subsection{Excitation from the Metastable State $5d^46s^2,{}^5D_4$}

The highest-$J$ quintet metastable, $5d^46s^2,{}^5D_4$ ($J_i=4$) exhibits a further redistribution of the dominant excitation channels. For $J_i=4$, a larger number of final states become dipole-allowed ($J_f=3,4,5$ from $\Delta J=\pm1,0$), and spin-orbit mixing continues to play a significant role as shown in Figs.~\ref{Figure6}(a-d). It is noticed from Fig.~\ref{Figure6}(a), the present RDW calculations show the strongest excitation from the $^5D_4$ state corresponds to transitions to the ($5d^46s(^6D)6p$) $^5D^{\circ}_{3}$, level with a peak cross-section of approximately $\sim$ 4.63$\times$ 10$^{-17}$ cm$^{2}$. This enhancement arises from the strong dipole coupling between the initial
$J_i$=4 state and final states with $J_f=3$.
 Along with that, transitions to $^7F^{\circ}_{5}$ (Fig.~\ref{Figure6}(a)) level of the $5d^46s(^6D)6p$ configuration reach peak cross-section of the order of 1.49 $\times$ $10^{-17}$~cm$^2$ and decreases slowly with increasing incident electron energy $E$. The excitations to $^7D^{\circ}_{3}$ and $^7D^{\circ}_{4}$ levels reach peak cross-sections on the order of 8.61 $\times$ $10^{-18}$~cm$^2$ but excitation to the $^7D^{\circ}_{4}$ decreases rapidly with increasing incident electron energy $E$ compared to $^7D^{\circ}_{3}$. Although these transitions involve septet states ($S = 3$), implying a nominal spin change from the initial quintet state ($S = 2$), they can be effectively excited through exchange coupling and strong configuration mixing, which are well-known features in tungsten. As shown in Figs.~\ref{Figure6}(a-d), the excitation cross-sections for the $^7F^{\circ}$, $^7P^{\circ}$, and $^5D^{\circ}$ manifolds increase with increasing total angular momentum $J$, indicating a preferential population of the higher-$J$ fine-structure levels. In contrast, other excitation channels from the $^5D_4$ state, such as transitions to the $^7F^{\circ}_{0,1,2}$ and $^5D^{\circ}_{0,1}$ levels, exhibit considerably smaller cross-sections, with some decreasing rapidly as the incident electron energy increases. Some of these weaker transitions are not plotted in the present figure due to their very low cross-section magnitudes. Overall, the results suggest that the higher-$J$ $^7D^{\circ}$, $^7F^{\circ}$, and $^5D^{\circ}$ final states are significantly populated from the $^5D_4$ level, whereas the lower $J$ members of these terms remain comparatively weak. In general, the total excitation cross-sections originating from the $^5D_4$ level are similar in magnitude to those from other quintet metastable states.

\subsection{Excitation from the Metastable State $5d^46s^2,{}^3P_0$}

\begin{figure}

    \begin{subfigure}[ht]{0.5\textwidth}
           \includegraphics[width=\textwidth]{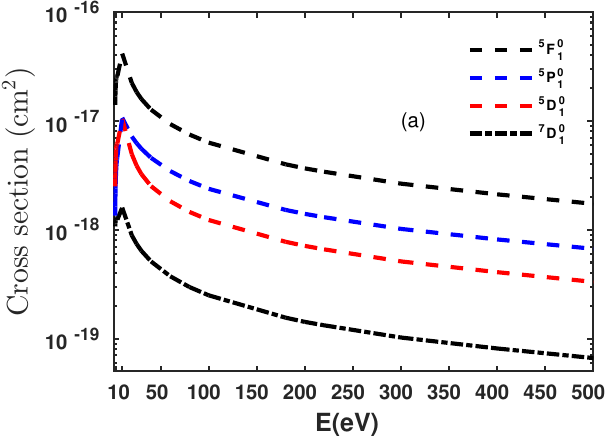}
     \end{subfigure}
     \hfill
     \begin{subfigure}[ht]{0.5\textwidth}
            \includegraphics[width=\textwidth]{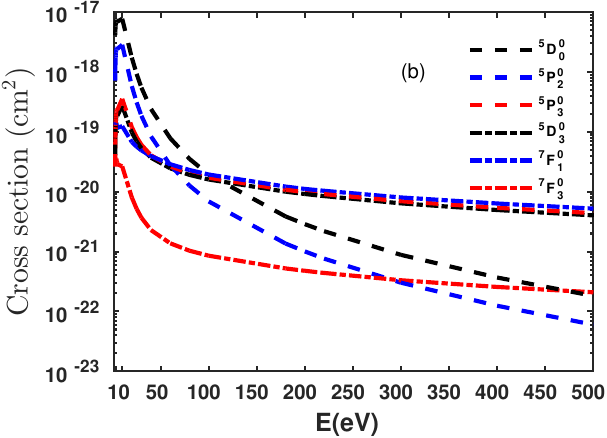}
     \end{subfigure}
\hfill
     \begin{subfigure}[ht]{0.5\textwidth}
           \includegraphics[width=\textwidth]{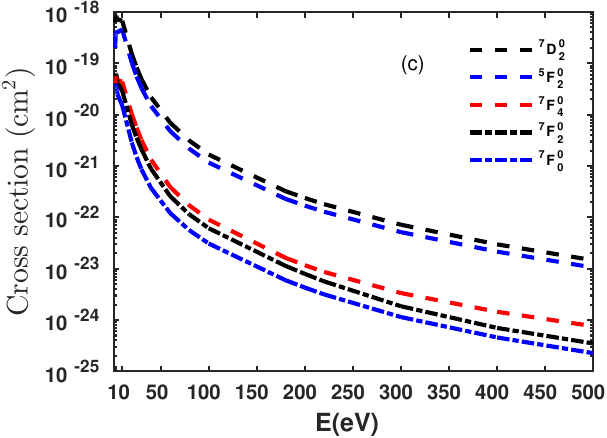}
     \end{subfigure}
\hfill
     \begin{subfigure}[ht]{0.5\textwidth}
           \includegraphics[width=\textwidth]{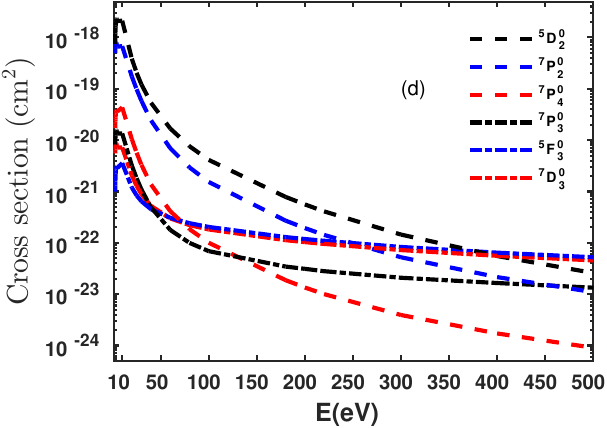}
     \end{subfigure}

\caption{Electron impact excitation cross-sections of W I from metastable state 5d$^4$6s$^2$ ($^3$P$_0$) to various excited states (a-d).}

\label{Figure7}

\end{figure}

Finally, the metastable $5d^46s^2,{}^3P_0$ ($J_i=0$, $S=1$) is considered in Figs.~\ref{Figure7}(a-d). This triplet state has a much lower total spin than the quintet and septet states discussed above, which significantly alters the coupling behavior. Overall, the absolute cross sections from $^3P_0$ are similar to other quintet states, but there remains a clear hierarchy among the various channels. Fig.~\ref{Figure7}(a) shows that the strongest transition from $^3P_0$ is found to be the excitation to the $5d^46s6p$ $^5F^{\circ}_1$ level. This channel benefits from exchange-driven coupling, and it reaches a peak of roughly 4$\times$$10^{-17}$~cm$^2$.
The $\sigma(^3P_0 \to {}^5F^{\circ}_1)$ transition also decreases slowly with $E$ compared to most other $^3P_0$ channels, allowing $^5F^{\circ}_1$ to retain a dominant influence at higher energies as well. The next most significant channels are the excitation to $^5D^{\circ}_1$, $^5P^{\circ}_1$  which peaks near $10^{-17}$~cm$^2$ and likewise shows a gradual decreasing tendency with energy $E$. These three ($^5F^{\circ}_1$, $^5P^{\circ}_1$ and $^5D^{\circ}_{1}$) clearly emerge as the primary excitation routes from the $^3P_0$ metastable. Apart from those, a septet channel ($^7D^{\circ}_1$) and two quintet channels $^5P^{\circ}_2$, $^5D^{\circ}_2$ achieve intermediate strength, the transition reach peak value on the order of $\sim$ 2-3 $\times$ $10^{-18}$~cm$^2$ as shown in Figs.~\ref{Figure7}(a) and \ref{Figure7}(b), respectively. The $^7D^{\circ}_1$ channel remains moderately strong over a fairly broad energy range, indicating that even from a triplet initial state, there is some ability, maybe through exchange and mixing, to excite higher-spin final states, though at a much reduced efficiency compared to a high-spin initial state. On the weaker side, most other transitions from $^3P_0$ are quite suppressed. For example, excitations to the $^7F^{\circ}_0$, $^7F^{\circ}_2$ and $^7F^{\circ}_4$ levels, as well as to $^7P^{\circ}_4$, as depicted in Figs.~\ref{Figure7}(c) and \ref{Figure7}(d), respectively. These channels involve $\Delta J=0 (0 \rightarrow 0)$, $2$, and $4$, which are forbidden for direct dipole and must proceed via higher multipoles and also couple a low-spin initial to specific high-spin final states, resulting in very small probabilities. In summary, the $^3P_0$ initial state primarily populates the quintet manifold especially the $^5P^{\circ}_1$, $^5D^{\circ}_1$ and $^5F^{\circ}_1$ terms. \par
Across all the initial states (ground $^5D_0$ and the six metastable levels $^5D_{1,2,3,4}$, $^7S_3$, and $^3P_0$), the calculated excitation cross sections exhibit the same basic physical behavior governed by selection rules. The largest cross sections in each case correspond to transitions that are electric-dipole allowed (opposite parity and $\Delta J=0,\pm1$) or only weakly forbidden, whereas transitions requiring higher-order multipoles or spin flips are smaller. The presence of numerous strong channels suggests that excitation from the $^7S_3$ state is broadly distributed over both $^7$ and $^5$ series levels, rather than being concentrated in a small subset of transitions.\par
\begin{figure}
\begin{centering}
           \includegraphics[width=0.8\textwidth]{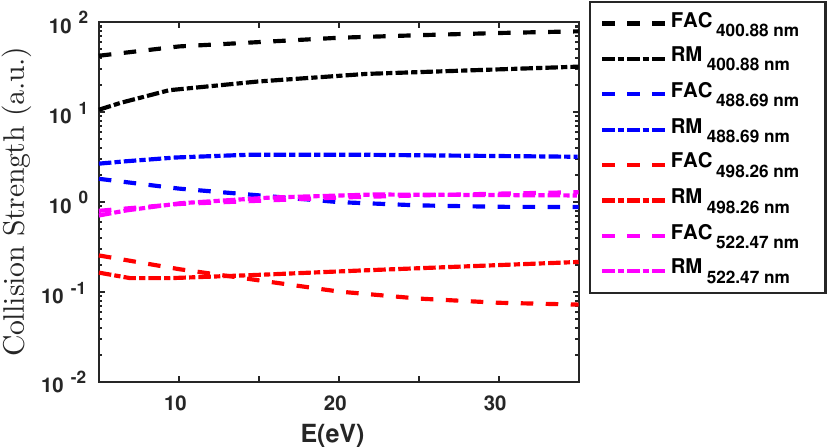}

\caption{The collision strengths calculated using FAC and the effective collision strengths obtained from the R-matrix calculations \cite{smyth} are shown by the dashed and dash-dotted lines, respectively, for the $5d^46s6p$ $^7P^{\circ}_4$-$5d^56s$ $^7S_3$ (400.88 nm), $5d^46s6p$ $^7F^{\circ}_5$-$5d^46s^2$ $^5D_4$ (488.69 nm), $5d^46s6p$ $^7F^{\circ}_1$-$5d^46s^2$ $^5D_0$ (498.26 nm), and $5d^46s6p$ $^7D^{\circ}_2$-$5d^46s^2$ $^5D_3$ (522.47 nm) transitions.}
\label{Figure8}
\end{centering}
\end{figure}

Furthermore, for selected transition wavelengths, the radiative transition probabilities have been evaluated and are presented in Table~\ref{Table2}. The radiative transition probabilities for the specific transitions are listed in Table ~\ref{Table2} along with the HFR \cite{hfr}, GRASP \cite {smyth}, MDFGME \cite {duck1} and the NIST \cite{Kramida} recommended values based on experiment. From Table~\ref{Table2}, it is observed that the radiative transition probabilities obtained from FAC show reasonably good agreement with the NIST, HFR, and MDFGME$_3$ results, often within the same order of magnitude for several transitions. While good agreement is obtained for several transitions with NIST, noticeable deviations remain for others. These deviations may arise from the strong configuration interaction and the highly complex atomic structure of neutral tungsten (W I). Therefore, further theoretical investigations, along with dedicated experimental measurements, are required to improve the reliability of the radiative transition probabilities for W I transitions.
 Furthermore, the calculated RDW cross-sections are presented up to an incident electron energy of 500 eV, whereas the collision strengths obtained from the R-matrix calculations \cite{smyth} are displayed up to 30 eV for four specific transitions: $5d^46s6p\ ^7P^{\circ}_4$ to $5d^56s\ ^7S_3$ (400.88 nm), $5d^46s6p\ ^7F^{\circ}_5$ to $5d^46s^2\ ^5D_4$ (488.69 nm), $5d^46s6p\ ^7F^{\circ}_1$ to $5d^46s^2\ ^5D_0$ (498.26 nm), and $5d^46s6p\ ^7D^{\circ}_2$ to $5d^46s^2\ ^5D_3$ (522.47 nm). For comparison, the collision strengths from FAC and R-matrix \cite{smyth} corresponding to these four transitions are plotted in Fig.~\ref{Figure8}. It is observed that the collision strengths from FAC differ from R-Matrix calculations for the transition wavelengths 400.88 nm, 488.69 nm and 498.26 nm. However, for the transition wavelength 522.47 nm, these two results exactly coincide with each other, shown by magenta lines in Fig.~\ref{Figure8}. The differences between the two sets of results can be attributed to the perturbative nature of the RDW theory, whose accuracy is limited near the threshold energy region.

\begin{table}[htbp]
\centering
\caption{Radiative transition rates ($s^{-1}$) of W I transitions from FAC, HFR \cite{hfr}, GRASP \cite {smyth}, MDFGME \cite {duck1} and the NIST \cite{Kramida} recommended values based on experiment.}

\resizebox{\textwidth}{!}{%

\begin{tabular}{lllllllll}
\hline
Wavel. & Trans. & \multicolumn{7}{l}{A$_{ij}$ ($s^{-1}$)} \\
 (nm)   &  j$\rightarrow$ i     & FAC & NIST & HFR & GRASP & MDFGME$_1$ & MDFGME$_2$ & MDFGME$_3$ \\
\hline
354.52 & 28-1 & 1.68E+06 & 3.20E+06 & 2.49E+06 &  & 6.44E+05 & 6.16E+05 & 7.85E+05  \\
376.84 & 28-2 & 3.88E+05 & 3.47E+06 & 3.08E+06 & 2.20E+07 & 2.93E+05 & 6.60E+05 & 1.01E+06 \\
381.74 & 20-3  & 3.64E+07 & 3.10E+06 & & & &\\

383.51 & 29-4  & 1.38E+05  & 5.20E+06 & 6.48E+06 & 2.63E+07 & 2.71E+06 & 1.74E+06 & 3.28E+06\\

384.62 & 19-2 & 2.31E+05 & 2.14E+06 & 2.12E+06 & 1.46E+07 & 5.74E+05 & 1.94E+03 & 4.54E+05  \\
386.79 & 17-3 & 1.28E+07 & 4.60E+06 &  &  &  &  &  \\
388.14 & 30-5 & 9.43E+04 & 3.60e+06 &  &  &  &  &  \\

400.87 & 23-3 & 6.82E+07  & 1.63E+07 & 1.83E+07  & 1.14E+07 & 2.54E+07 & 2.34E+07  & 2.41E+07  \\

404.56 & 19-3 & 2.64E+07 & 2.88E+06  & 2.78E+06  &  & 4.32E+05 & 1.76E+05  & 3.41E+05 \\

 407.44 & 22-3 & 4.19E+07 & 1.00E+07 & 9.71E+06 & 5.66E+06 & 1.82E+07 & 1.60E+07 & 1.82E+07  \\

410.27 & 30-6  & 8.11E+04 & 4.90E+06 & 4.54E+06 & 5.39E+06  & 6.63E+05 & 4.40E+06 & 4.51E+06 \\
413.75 & 22-4 & 1.21E+06 & 8.40E+05 &  &  &  &  &  \\

429.46 & 29-3 & 3.08E+07 & 1.24E+07 & 1.22E+07  & 3.17E+07 & 2.45E+07 & 2.14E+07 & 2.37E+07  \\

430.21 & 16-3 & 2.76E+06 & 3.60E+06  & 4.04E+06 & 4.56E+06 & 9.06E+06 & 8.68E+06 & 8.11E+06\\

465.99 & 14-1 & 8.39E+05 & 1.00E+06 & 1.24E+06 &  & 9.28E+05  & 1.07E+06  & 1.01E+06\\

468.05 & 16-5 & 4.64E+05  & 1.40E+06 & 2.40E+06 &  & 4.82E+06 & 7.02E+05 & 7.23E+05 \\

484.38 & 15-4 & 1.04E+06 & 1.90E+06 & 2.92E+06 &  & 1.06E+06 & 1.29E+06 & 1.28E+06 \\

488.69 & 13-6 & 2.11E+05  & 8.10E+05 & 1.69E+06 & 1.48E+06   & 2.20E+05 & 3.01E+05 & 3.13E+05 \\

498.26 & 9-1 & 4.82E+04  & 4.17E+05 & 1.07E+06 & 2.13E+05 & 5.53E+04 & 8.84E+04 & 1.05E+05 \\

500.61 & 16-6 & 5.11E+05 & 1.20E+06 & 1.66E+06 & & 5.02E+05  & 6.88E+05 & 6.87E+05 \\

505.33 & 14-2 & 1.36E+06 & 1.90E+06 & 2.51E+06 & 3.24E+06 & 1.35E+06 & 1.70E+06 & 1.64E+06 \\

522.47 & 15-5 & 7.45E+05 & 1.20E+06 & 1.64E+06 & 1.23E+06  & 8.34E+04  & 9.44E+05 & 9.04E+05 \\
535.44 & 28-7 & 1.45E+05 & 1.00E+05 & & & & & \\
539.10 & 12-6 & 3.41E+03 & 8.90E+03 & & & & & \\

551.47 & 14-4  & 5.75E+05 & 7.30E+05  & 9.97E+05 & & 5.21E+05 & 7.29E+05 & 7.04E+05    \\ \hline

\end{tabular}}
\label{Table2}
\end{table}

\newpage
\section{Conclusions}
In conclusion, fine-structure-resolved electron-impact excitation cross sections for a wide range of transitions in neutral tungsten (W I) have been systematically calculated using the relativistic distorted-wave method within the FAC framework. The inclusion of an extensive configuration set ensures a proper representation of both valence–valence and core–valence correlation effects in the present calculations. The calculated energy levels have been compared with the HFR, GRASP, and MDFGME results, as well as with the recommended values from the National Institute of Standards and Technology (NIST) database. The lower-lying energy levels obtained in the present work show good agreement with the NIST data. The application of an additional low-energy correction enables reliable description of the cross sections in the near-threshold energy region. The cross sections have been computed over an incident electron energy range of 0-500 eV. Excitations originating from the ground state $5d^46s^2\ (^{5}D_0)$ and selected metastable states, namely $5d^46s^2\ (^{5}D_{1,2,3,4},\ ^3P_0)$ and $5d^5(^{6}S)6s\ (^{7}S_3)$, to the excited configurations $5d^46s(^{6}D)6p$ and $5d^56p$ exhibit a strong dependence on the initial state. The results indicate that excitation from the $5d^5 6s\ (^{7}S_3)$ metastable state produces the largest cross sections, whereas transitions originating from the $5d^46s^2$ levels show comparable magnitudes. These findings confirm that metastable states make a significant contribution to the population of the excited levels. In addition, radiative transition probabilities for several prominent W I transitions have been evaluated and compared with the available literature data. The observed differences between the present FAC results and other theoretical values can be attributed to the complex configuration interaction effects in neutral tungsten. Overall, the present work provides a comprehensive and self-consistent dataset that is expected to be valuable for plasma modelling and spectroscopic diagnostics involving tungsten. Future work will involve incorporating the present atomic dataset into a collisional–radiative framework to investigate the role of metastable-state populations in determining excited-level populations and the resulting spectral line intensities over a wide range of plasma conditions.

\section{Acknowledgements}
The authors acknowledge Board of Research in Nuclear Sciences (BRNS), DAE, Government of India, for supporting this work under research project grant (sanction no. 57/14/01/2023/12089). The simulation results presented in this study were obtained using ANTYA, the IPR Linux Cluster.
\newpage

\textbf{References}\\

\end{document}